\documentstyle[prl,aps,twocolumn,multicol,epsf]{revtex}
\tighten
\begin{document}
\twocolumn[\hsize\textwidth\columnwidth\hsize\csname @twocolumnfalse\endcsname
\title {Late-time evolution of nonlinear gravitational collapse}
\author{Lior M. Burko and Amos Ori}
\address{
Department of Physics, Technion---Israel Institute of Technology,
32000 Haifa, Israel.}
\date{\today}



\maketitle

\begin{abstract}

We study numerically the fully nonlinear
gravitational collapse of a self-gravitating, minimally-coupled,
massless scalar field in spherical symmetry. Our numerical code is based
on double-null coordinates and
on free evolution of the metric functions: The evolution equations
are integrated numerically, whereas the constraint equations are
only monitored. The numerical code is stable (unlike recent claims) 
and second-order accurate. 
We use this code to study the late-time asymptotic behavior
at fixed $r$ (outside the black hole), along the event horizon, and along
future null infinity.
In all three asymptotic regions we find that, after the decay of
the quasi-normal modes, the perturbations are dominated by
inverse power-law tails.
The corresponding power indices agree with the integer
values predicted by linearized theory.
We also study the case of a charged black hole nonlinearly
perturbed by a (neutral)
self-gravitating scalar field, and find the same type of behavior---i.e., 
quasi-normal modes followed by inverse power-law tails,
with the same indices as in the uncharged case.
\newline
\newline
PACS number(s): 04.70.Bw, 04.25.Dm
\end{abstract}

\vspace{3ex}
]

\section{Introduction}
The no-hair theorems state that except for the mass, the
electric charge, and the angular momentum, all the features of
fields which collapse to a black hole will be unobservable
to external observers at late times.
It is therefore interesting to study the
mechanism by which the hair
is radiated away (or absorbed by the black hole).

Until recently, the late-time evolution of
non-spherical gravitational collapse was
investigated primarily in the context of linear theory.
That is, the deviations
from spherical symmetry were considered as 
infinitesimally-small perturbations over a fixed
curved background. The late-time behavior of such perturbations has been
studied for three different asymptotic regions:
(a) at fixed $r$, (b) along null infinity, and
(c) along the future event horizon (when the collapse is to a black hole).
Qualitatively, the evolution of the linearized perturbations
is similar in these three asymptotic regions: During the first stage,
the perturbations' shape depends strongly on the shape of the initial
data. This stage is followed by the stage of
quasi-normal (QN) ringing,
in which the perturbations oscillate with an exponentially-decaying
amplitude. The corresponding complex frequency is characteristic of the
parameters of the background
black-hole, and is independent of the details of the initial perturbation.
Finally, there are also `tails',
characterized by an inverse power-law decay.

The asymptotic region (a) was first studied by Price \cite{price},
who analyzed the linear perturbations over
a fixed Schwarzschild background. Price found that after the QN
ringings die out, the perturbations at fixed $r$ (outside the black hole)
decay according to $t^{-(2l+\mu+1)}$, where $\mu=1$ if there were an
initial static mode, and $\mu=2$ otherwise. Here, $l$
is the multipole moment of the mode in question, and $t$ is the
standard external Schwarzschild time coordinate.
Asymptotic regions (b) and (c) were considered by
Gundlach, Price and Pullin \cite{GPP1},
who showed that the
tails along null infinity decay according to $u_e^{-(l+\mu)}$,
where $u_e$ is the outgoing Eddington-Finkelstein null coordinate.
(Hereafter we use the notation $u_e$ and $v_e$ for the outgoing
and ingoing Eddington-Finkelstein
null coordinates, correspondingly, in order to
distinguish them from other types of null coordinates which we use
later.) Along the
event horizon (EH), the inverse-power indices were found to be similar to
the asymptotic limit (a); Namely, the `tails' decay according to
$v_e^{-(2l+\mu+1)}$.

Recently, Krivan, Laguna and Papadopoulos studied
numerically the evolution of linearized scalar-field \cite{KLP} 
and spin-2 perturbations 
\cite{KLP1} over a fixed Kerr background. (See also \cite{KLP2}.)
They concluded that `tails' are expected also for the Kerr
background, with power-law indices similar to those obtained for
Schwarzschild. This provides additional motivation for
the study of the fully-nonlinear evolution
of perturbations in the spherically-symmetric case as a toy-model for
the spinning case, since the spherical case is much simpler to deal
with (both analytically and numerically).

The numerical simulation of the fully-nonlinear
gravitational collapse of a spherically-symmetric self-gravitating
scalar field was recently carried out by two groups: Gundlach, Price
and Pullin \cite{GPP2} (GPP), and Marsa and Choptuik (MC) \cite{MC}.
In both analyses,
the coordinates used were non-vacuum generalizations of the
(one null+$r$) outgoing \cite{GPP2} or ingoing \cite {MC} 
Eddington-Finkelstein coordinates.
These numerical analyses demonstrated the QN ringing
as well as the power-law `tails' for lines $r={\rm const}$. In addition, MC also
demonstrated the power-law decay at the EH.

In this paper, too,            
we study the nonlinear spherical gravitational collapse                  
of a self-gravitating scalar field. However,                      
we shall use different coordinates,                  
different numerical methods, and a somewhat different model.
Our numerical code is stable and second-order accurate and 
is based on free evolution and double-null coordinates. 
This combination has several advantages:
First, the null coordinates are       
very well adapted to the hyperbolic character of the field equations
evolved:
The evolution is along the characteristics, and consequently there
is no restriction analogous to the Courant condition.
Second, in double-null coordinates the
interpretation of the causal structure of
the numerically-produced spacetime                            
is trivial. Also, double-null coordinates                    
can be chosen such that the metric is
regular at the EH, which is not the case in outgoing 
Eddington-Finkelstein coordinates. 

Our coordinates and integration scheme
allow us to study the evolution to arbitrarily late times,
and there is no need to introduce an artificial outer boundary \cite {MC}.
Our analysis demonstrates both the QN ringing and the power-law tails.
One of our main objectives in this investigation
is to numerically determine the power-law indices of the late-time tails
in the nonlinear collapse problem, and to compare them to the predictions
of the linear perturbation theory.                                        
In cases (a) and (c), we obtained power-law indices similar to
those found by \cite{GPP2,MC}, though with improved accuracy.
In addition, we                                                              
obtain the power-law index at null infinity, which was not studied
so far in nonlinear collapse.                       
In all three asymptotic limits, we find an excellent agreement between our
numerically-obtained indices and the values
predicted by the linear perturbation
analyses \cite {price,GPP1}. (Such an agreement is expected, even in a
very
nonlinear collapse problem, because of the ``no-hair'' principle---see
e.g. Ref. \cite {GPP2}.)        

Whereas this paper considers only the external part of the black hole,
we are currently investigating the inner structure of charged black holes
with a similar numerical code \cite{remark}. 
In fact, our main motivation in this project
is to develop the numerical approach and techniques which could later be
used in investigating the black hole's interior. The determination of
the correct late-time power-law index is essential for that purpose.

Similar self-gravitating collapse scenarios have been recently used for the 
study of critical phenomena in black hole formation 
\cite{choptuik93,HS,gundlach97}. At its present form our code is incapable of 
treating these phenomena because we have not attempted to include the neighborhood 
of the origin in the domain of integration. The configurations we are interested in here 
are by far super-critical, and the aspects which concern us do not require the 
integration near the origin (see below). 

This paper is organized as follows: In Section II we present the model
for the collapse, and the corresponding field equations.
Section III describes the numerical approach, and Section IV discusses
the stability and accuracy of our code, and the tests used to verify them.  
It has been recently argued 
\cite{GP} that 
unconstrained codes in double-null coordinates suffer from inherent instabilities. 
We show that this is not the case, and that our code is indeed stable and 
converges with second order. (In the Appendix we explain this in greater 
detail.) In Section V we present our numerical results for
the collapse of a scalar field over a Minkowski background, leading to
the formation of a Schwarzschild black hole, and in
Section VI we consider the collapse of a (self-gravitating, neutral)
scalar field on a RN background. Finally, in section VII
we summarize and discuss our results.

\section{The Collapse Model}
\subsection{The field equations}
We shall consider the spherically-symmetric
gravitational collapse of a self-gravitating, minimally-coupled,
massless scalar field. In
the uncharged case, the system is described by the
coupled Einstein-Klein-Gordon field equations. We shall also consider
the charged case, i.e. the case in which a
(source-less) spherically-symmetric
electric field is also present. In this case,
the system is described by the
coupled Einstein-Maxwell-Klein-Gordon field equations.

We write the field equations
in double-null coordinates. The line element takes the form
\begin{eqnarray}
\,ds ^{2}=-f(u,v)\,du\,dv+r^{2}(u,v)\,d\Omega^{2},
\label{metric}
\end{eqnarray}
where $\,d\Omega^{2}$ is the line element on the unit two-sphere.
The general spherically-symmetric solution of the Maxwell equations
in these coordinates is 
\begin{eqnarray}
F_{uv}=-F_{vu}=\frac{1}{2}\frac{Qf}{r^{2}}
\end{eqnarray}
and $F_{\mu \nu}=0$ otherwise, where Q is a free parameter,
which is interpreted as the electric charge, and where $F_{\mu\nu}$ 
is the Maxwell field tensor.  
The contribution of this electric field to the energy-momentum tensor is
\begin{eqnarray}
T_{\mu\nu}^{\rm em}=\frac{Q^{2}}{8\pi r^{4}}
\left(\begin{array}{cccc}
0 & f/2 &0 & 0 \\
f/2 &0 &0 &0 \\
0 &0 &r^{2} &0 \\
0 &0 &0 &r^{2}\sin^{2}\theta 
\end{array} \right) . 
\end{eqnarray}
The energy-momentum tensor of a massless scalar field $\Phi$ is
\begin{eqnarray}
T_{\mu\nu}^{\rm s}=\frac{1}{4\pi}\left(\Phi_{,\mu}\Phi_{,\nu}-\frac{1}{2}
g_{\mu\nu}g^{\alpha\beta}\Phi_{,\alpha}\Phi_{,\beta}\right).
\end{eqnarray}
This field satisfies the Klein-Gordon equation ${\Phi_{;\alpha}}^{;\alpha}=0$,
which, in our coordinates, takes the form
\begin{eqnarray}
\Phi_{,uv}+\frac{1}{r}\left(r_{,u}\Phi_{,v}+r_{,v}\Phi_{,u}\right)=0.
\label{KGEQ}
\end{eqnarray}

The Einstein field equations are
$G_{\mu\nu}=8\pi T_{\mu\nu}$, where the
energy-momentum tensor is the sum of the contributions of both
the electromagnetic and scalar fields,  
$T_{\mu\nu}=T_{\mu\nu}^{\rm s}+T_{\mu\nu}^{\rm em}$. These
equations include two evolution equations,
\begin{eqnarray}
r_{,uv}=-\frac{r_{,u}r_{,v}}{r}-\frac{f}{4r}\left(1-\frac{Q^{2}}
{r^{2}}\right)
\label{EEQ1}
\end{eqnarray}
and
\begin{eqnarray}
f_{,uv}&=&\frac{f_{,u}f_{,v}}{f}+
f\left\{ \frac{1}{2r^{2}}\left[4r_{,u}r_{,v}\right.\right.\nonumber \\
&+&\left.\left.
f\left( 1-2\frac{Q^{2}}{r^{2}}\right)\right]-
2\Phi_{,u}\Phi_{,v} \right\} \; ,
\label{EEQ2}
\end{eqnarray}
supplemented by two constraint equations:
\begin{eqnarray}
r_{,uu}-(\ln f)_{,u}r_{,u}+r(\Phi_{,u})^{2}=0
\label{con1}
\end{eqnarray}
\begin{eqnarray}
r_{,vv}-(\ln f)_{,v}r_{,v}+r(\Phi_{,v})^{2}=0.
\label{con2}
\end{eqnarray}
The constraint equations are not independent of the dynamical equations:
Any solution of the evolution equations will also be a solution
of the constraint equations, provided only that the latter are satisfied
on the initial hypersurface. (This is assured 
by virtue of the contracted Bianchi indentities.) 

\subsection{The formulation of the characteristic problem}

In our numerical scheme, we shall use the three equations
(\ref{KGEQ})--(\ref{EEQ2}) to evolve the three unknowns
$r(u,v)$, $f(u,v)$, and $\Phi(u,v)$. These equations form a
hyperbolic system, and thus ensure a well-posed initial-value
formulation. In the double-null coordinates we use, it is most
natural to use the characteristic initial-value formulation,
in which the initial values of the unknowns (but not of
their derivatives) are specified on two null segments,
$u={\rm const}\equiv u_i$ and $v={\rm const}\equiv v_i$.

In such a numerical scheme (often called {\it free evolution}),
the constraint equations are only
imposed on the initial hypersurfaces. As mentioned above, the
consistency of the evolving fields with the constraint equations
is mathematically guaranteed. We use the constraint equations to
check the accuracy of the numerical simulation.

From the pure initial-value viewpoint, we need to specify three
initial functions on
each segment of the initial surface: $r$, $f$, and $\Phi$.
The constraint equations, however, reduce
this number by one: Eq. (\ref{con1})
imposes one constraint on the initial data at $v=v_i$, and similarly,
Eq. (\ref{con2})
imposes one constraint on the initial data at $u=u_i$. The
remaining two initial functions,
however, represent only {\it one} physical degree of freedom: The other
degree of freedom expresses nothing but the gauge freedom
associated with the arbitrary coordinate transformation
$u\to \tilde u(u)\;,\;v\to \tilde v(v)$
(the line element (\ref{metric}) and all the above equations are invariant
to this transformation).
In what follows we shall use a standard
gauge, in which $r$ is linear with $v$ or $u$, correspondingly,
on the two initial null segments. On
the outgoing segment we take $r_{,v}=1$. On the ingoing segment,
we take $r_{,u}={\rm const}\equiv r_{u0}$.
The initial values of $r$ are thus uniquely
determined by the parameter $r_0\equiv r(u_i,v_i)$. 
We choose $u_{i}=0$ and $v_{i}=r_{0}$, and thus we find:
$$r_v(v)=v \;\;,\;\; r_u(u)=r_0+u r_{u0}\;.$$
(Hereafter, we denote the initial values of the three fields on
the two initial segments by $r_u(u),f_u(u),\Phi_u(u)$ and
$r_v(v),f_v(v),\Phi_v(v)$, correspondingly.) 
Then, we can freely specify $\Phi_u(u)$ and $\Phi_v(v)$
(this choice represents a true physical degree of freedom).
The initial value of $f$ is now
determined from the constraint equations, namely
\begin{eqnarray}
(\ln f_u)_{,u}=r_u(\Phi_{u,u})^{2}/r_{u0}\;\;,\;\;
(\ln f_v)_{,v}=r_v(\Phi_{v,v})^{2}\;,
\label{initf}
\end{eqnarray}
together with the choice $f(u_i,v_i)=1$.
Thus, in the gauge we use, the geometry in the entire
domain of dependence is uniquely determined by the two initial functions
$\Phi_u(u)$ and $\Phi_v(v)$, and the two parameters
$r_0$ and $r_{u0}$. (Later we shall relate $r_{u0}$ to the initial
black-hole mass.)
In what follows, we shall consider initial data corresponding to a compact
ingoing scalar-field pulse, over a background of either Minkowski,
Schwarzschild, or RN.
Namely, we shall assume that $\Phi_u(u)\equiv0$; and
$\Phi_v$ is also zero, except at a finite interval $v_1<v<v_2$
(with some $v_1 \ge v_i$) where $\Phi_v\ne 0$.
For concreteness, in the range $v_1<v<v_2$ we shall take 
$\Phi_v=A\sin ^{2}[\pi(v-v_{1})/(v_{2}-v_{1})]\;$.  
This choice for the initial data is smooth at the matching points 
$v=v_1$ and $v=v_2$. 
The determination of $f_{v}(v)$ throughout $u=u_{i}$, by analytically 
integrating the second constraint equation in (\ref{initf}), 
is straightforward for such a pulse. On 
the ingoing segmant $v=v_{i}$ we have $f_{u}(u)=1$.   

The geometry is static (with $\Phi=0$) in the entire
range $v<v_1$, with a mass parameter $M_0$.
To relate $M_0$ to the above initial-value parameters, we
define the mass-function $M(u,v)$ \cite {Poisson-Israel} by
$r_{,\mu} r^{,\mu} =1-2M(u,v)/r+Q^2/r^2$.
In our coordinates this becomes
\begin{eqnarray}
M(u,v)=(r/2)(1+4 f^{-1} r_{,u} r_{,v})+Q^2/2r\;,
\label{mass-function}
\end{eqnarray}
which yields $M_0=(r_0/2)(1+4 r_{u0})+Q^2/2r_0\;.$
Thus, our initial-value set-up is determined by the initial
mass-parameter
$M_0$, the charge $Q$, and the perturbation amplitude $A$
(together with the auxiliary parameters $v_1$, $v_2$ and $r_0$).

We shall particularly study two cases:
(i) $M_0=0,Q=0$ (Minkowski background),
and (ii)  $M_0=1,Q\ne 0$ (RN background).
[Note that no loss of generality is caused by the choice
$M_0=1$, because of the scale-invariance nature of the problem.]
We shall not elaborate here on the
situation of self-gravitating scalar field collapsing over a
Schwarzschild background, as its outcome
is qualitatively similar to case (i)
(and, the collapse over Minkowski brings out the nonlinear aspects
in a sharper way). We shall use, however, the pure ($A=0$) Schwarzschild
case as a test-bed for our numerical code.

In what follows, we shall use the symbols $u$ and $v$ to denote
the outgoing and ingoing null coordinates
in the specific gauge described above.
Notice that $v$ is closely related to $v_e$ at $v\gg M$, and $u$ is
Kruskal-like near the EH (namely, it regularizes
the metric function $f$ at the EH).

The double-null line element suffers from a non-physical coordinate singularity
at the origin (i.e., the timelike worldline $r=0$, where the geometry
is perfectly regular). This singularity may cause difficulties
in the numerical study of case (i) above (i.e., Minkowski background).
In order to overcome this
difficulty, we restrict the domain of numerical integration
in this case such that it will not include the origin. That is, the
characteristic initial segment $v=v_i$ ends before it reaches $r=0$.
Since it is very essential that the domain of integration will include
the EH, we must demand that the ingoing ray $v=v_i$
will intersect the EH before it intersects $r=0$. This is
achieved if the amplitude parameter $A$ is sufficiently large.

\section{The Numerical code}

Our numerical code is based on the standard procedure for
second-order integration of (1+1) hyperbolic equations in double-null
coordinates: Let $\,du$ and $\,dv$ be the finite increments in the $u$ and
$v$ directions, respectively. Let us also denote
schematically the three unknowns $r,f,\Phi$
as $h_i\;,\;i=1,3$; These unknowns satisfy a field equation of the form 
\begin{eqnarray}
h_{i,uv}=F_i(h_j,h_{j,u},h_{j,v})\;,\;j=1,3\; \; .
\end{eqnarray}
Assume now that we already know the values of $h_i$ at
the three grid points $p_1\equiv (u_0,v_0)$, $p_2\equiv (u_0+\,du,v_0)$, and
$p_3\equiv (u_0,v_0+\,dv)$, and we would like to evaluate them at
$p_4\equiv (u_0+\,du,v_0+\,dv)$. We then use the substitution
\begin{eqnarray}
h_i^{(4)}=-h_i^{(1)}+h_i^{(2)}+ h_i^{(3)}+F_i^{(5)} \,du\,dv \; \; ,
\label{fourpoints}
\end{eqnarray}
where, for any function $g$, $g^{(k)}\equiv g(p_k)$, and
$p_5$ is the intermediate point: $p_5\equiv (u_0+\,du/2,v_0+\,dv/2)$.
In order to evaluate the functions $h_j^{(5)},h_{j,u}^{(5)},
h_{j,v}^{(5)}$ (required for the determination of $F_i^{(5)}$)
to the desired accuracy, we use the standard ``predictor-corrector'' 
method. This procedure results in a second-order accuracy. With this method,
we first calculate $h_i$ along the ingoing ray $v=v_i+\,dv$---starting at
$u=u_i+\,du$, then solving for $u=u_i+2\,du$, and so on,  
until the last grid point at $u=u_f$;
Then we turn to the next ingoing ray $v=v_i+2\,dv$, and solve for all grid
points along this ray; By this way we solve,
ray after ray, until we cover the
entire domain of integration, $v_i<v<v_f \; , \; u_i<u<u_f$.
The accuracy is controlled by the global
grid parameter $N$, which is the (initial)
number of points per a unit interval
in both the $v$ and $u$ directions. Typically we used $N=10,20$ or $40$,
though in certain cases we also used the values $5$, $80$, and $160$.

As long as our domain of numerical integration does not include the EH, 
this numerical scheme can be used with a fixed grid without any
difficulties. When the EH is included,
however, we face a fundamental difficulty.
For simplicity, assume at this stage that $\Phi=0$ and the
spacetime is Schwarzschild (though the same conceptual
difficulty arises also in non-static spacetimes).
Let us denote by $u_h$ the value of our Kruskal-like coordinate
$u$ at the EH. 
Let $u_0$ be a grid value of $u$ just near the EH,
and let $u_1$ be the next grid point in $u$, i.e. $u_1\equiv u_0+\,du$.
We define $\delta r(v)\equiv r(u_1,v)-r(u_0,v)$.
Of course, for the validity of the numerical integration
it is necessary that $\delta r\ll r$ --- and we shall indeed select the grid
parameter $\,du$ sufficiently small so as to satisfy this requirement at the
initial segment $v=v_i$. The problem is that, $\delta r$ grows unboundedly
and very rapidly with $v$. Thus, in terms of
the Eddington-Finkelstein coordinate $v_e$, along the horizon
$\delta r\propto \exp(v_e/4M)$
(because at the horizon of Schwarzschild, $r_{,u}\propto \exp(v_e/4M)$, 
and $\delta r(v)\cong r_{,u}(u_0,v)\,du$.)
It is therefore obvious that a code based on a fixed $\,du$ cannot be used
here. One might attempt to use a numerical scheme in
which $\,du$ depends on $u$ (but not on $v$)
in such a way that it becomes extremely small at the EH. But this
turns out to be impractical too, because of the extremely large exponential
factor: Typically we need to integrate up to $v_e$
values of at least a
few hundreds times $M$ (otherwise we cannot study the power-law tails with
a sufficient accuracy). This would demand a value of $\,du$ as small as, say,
$10^{-100}$ near the EH, which is obviously impractical, due
to the roundoff error and other reasons.

In order to overcome this difficulty, we must use a dynamical 
grid-refinement algorithm. A sophisticated dynamical refinement scheme was 
recently
developed by Hamad\'{e} and Stewart \cite{HS}, in order to analyze the 
critical
behavior at the origin. For our purposes, however, it is sufficient to
use a simpler refinement scheme, which we call {\it point 
splitting}: In certain values of $v$, we check the variations in $r$
(and, in fact, in all $h_i$) between any two adjacent grid points. If the
difference in $r$ between such two points $p_1=(u_0,v_0)$
and $p_2\equiv (u_0+\,du,v_0)$ is
greater than some threshold value, we introduce an intermediate grid point
$p'_2\equiv (u_0+\,du/2,v_0)$,
and calculate the interpolated values of all unknowns
$h_i$ at that point. We can now use the above three-points
integration scheme to calculate $h_i$ at $p'_4\equiv (u_0+\,du/2,v_0+\,dv)$
from the values of these fields at $p_1,p'_2,p_3$ (and, later, to calculate
$h_i$ at $p_4\equiv(u_0+\,du,v_0+\,du)$ according to
the field values at $p'_2,p_2,p_4'$).
This numerical procedure functions very well in double-null coordinates,
especially due to the following reasons:
First, in the three-points integration
scheme, $(p_1,p_2,p_3)\to p_4$, there is no reference to any grid points at
$u<u_0$ or $u>u_0 +\,du$. Therefore, it does not matter whether 
the increments $\,du$ are uniform or not. Second, in this scheme there is no
restrictions on the ratio of $\,du$ and $\,dv$.

In practice, we proceed as follows. We register all grid values
of $u$ (at a given $v$) in a vector $u(I)$, where $I$ is an integer index. 
The values of the
three unknowns are registered in corresponding three vectors $h_i(I)$.
We define three threshold parameters $h_i^c$ for the three unknowns,
and also a ``band parameter'' $v_b$                                         
(typically we take $v_b$ to be of order $M$).                     
At the end of each interval $v_b$ in $v$ (a ``band''),                    
we check the variation of all three fields along                          
the vectors $h_i(I)$.                                                     
If for a given $I$ the relative difference                    
$\left| {[h_i(I+1)-h_i(I)]/h_i(I)} \right|$ is found to be   
greater than $h_i^c$ (for any $i$), then we add a                            
new grid point at $u=[u(I)+u(I+1)]/2$.                            
In such a case, we calculate the values of the fields
$h_i$ at that new point by interpolation (usually we perform a four-points
interpolation, based on, e.g., $u(I-1),u(I),u(I+1),u(I+2)$).
We now update the vectors $u(I)$ and $h_i(I)$,                
by assigning the value $I+1$ to the new grid point.                        
(Before creating this new grid point, we arrange an empty ``slot'' for it,
by shifting all grid index values $I'>I$ by one.)                      
The threshold values $h_i^c$ are taken to be proportional to $1/N$, in
order to preserve the rule of $N$ as a parameter that controls the global
accuracy (that is, the number of grid points in the $u$ axis should be
proportional to $N$). The band parameter $v_b$ is taken to be independent
of $N$.

Because our goal is to study the evolution in the entire                
black-hole exterior,                                                     
the domain of integration must include the EH and thus     
extend into the black hole (i.e., $u_f>u_h$).                             
Then, if $Q=0$, the numerical                                     
integration will terminate at some finite $v$, beyond which   
the ingoing null lines $v={\rm const}$ intersect the                            
spacelike $r=0$ singularity (before $u=u_f$). In order to overcome
this difficulty, 
we simply chop the vector $u(I)$ just beyond the apparent
horizon (AH). Namely, at the end of each band, we first find $I_{AH}$,
the value of the index $I$ where the AH is located.
This is the value of $I$ satisfying
$r_{,v}(u=u(I-1))>0$ and $r_{,v}(u=u(I))<0$.\footnotemark \footnotetext{
We also calculate $u_{AH}$ by 
interpolating between the two points $u(I-1)$ and $u(I)$. 
(In the Schwarzschild or RN cases $u_{AH}\equiv u_h$, but 
in the general dynamic case $u_{AH}\ge u_h$. 
Recall that only the AH can 
be found locally. However, for large $v$ the 
AH should coincide with the EH.)}
We then chop the vector
$u(I)$ at, say, $I=I_{AH}+1$. This ensures that the domain of integration never
gets close to the spacelike $r=0$ singularity---and yet it contains the
entire external part of the domain $v_i<v<v_f\;,\;u_i<u<u_f$, up to (and 
including) the EH. 

With these procedures of point-splitting and chopping, our code can
in principle run to arbitrarily large $v$ values. Due to point-splitting,
however, the number of grid points in the vector $u(I)$ grows 
linearly with $v_e$,
so the integration time grows like ${v_{e}}^{2}$. In order to significantly
decrease this time,  we introduce two additional types of numerical
manipulations:

{\it point removal:}
The successive addition of points near the event
horizon results in a coverage of the regions $r\gg 2M,v_e\gg M$ with an
approximately uniform density of order $N$ points per unit $u_e$
(Recall that a point added
at $u$ just before the EH will, after an interval $v_e\gg M$,
approach $r\gg 2M$.) However, for an appropriate coverage of
the variations in $r$, a much smaller
density of about $N M/r$ points per unit $u_e$
will be sufficient (note that at $r\gg 2M$, $r$ is approximately
linear in $u_e$ along lines $v={\rm const}$).
Thus, in order to save integration time, at the end of each
band we check along the vector $u(I)$ and simply remove all unnecessary points,
thereby shortening this vector. The criterion for necessity or otherwise of
a point $I$ is qualitatively similar to that of point-splitting:
Again, we define the threshold values ${h'}_{i}^{c}$ for point removal
(typically, ${h'}_{i}^{c}$ is slightly smaller than $h_{i}^c/2$). If for all
$i$ we have $\left| {[h_i(I-1)-h_i(I)]/h_i(I)} \right|<{h'}_{i}^{c}$,
then the point $I$ is removed\footnotemark
\footnotetext{In fact, the criterion we use
for the variation in $\Phi$ (for both
point-splitting and point removal) is somewhat more involved: It refers
to the variations in both $\Phi$ and $\Phi_{,u}$. This is essential for
the appropriate coverage of the maxima and minima 
regions in the QN ringing.}.

{\it gauge correction:}
As it turns out, for any Kruskal-like $u$ (which is necessary for a 
regular coverage of the EH) and Eddington-like $v$, $f$ grows
exponentially with $v_{e}$ along the EH. In addition,
along lines $v={\rm const}\gg M$, $f$ grows rapidly (exponentially
with $u_e$) in most of the interval $u_i<u<u_h$.
The above numerical scheme handles very well this behavior of $f$. However,
the significant variation of $f$ with $u$ implies that points can hardly
be removed, which results in a long computation time. 
In order to overcome this difficulty, we
introduce (as an option) a gauge correction at the
end of each ``band.'' That is, we perform a coordinate transformation
$u\to u_{new}(u)$. The value of $f$ is gauged
accordingly: $f_{new}=(\,du_{old}/\,du_{new}) f_{old}$
(the variable $r$ is unchanged). Our field equations
are invariant to such a coordinate transformation.
The function $u_{new}(u)$ is to be chosen so as to
decrease the variation of $f_{new}$ with $u$. A convenient choice is to
take $u_{new}(u,v_0)=r(u,v_0)$ (where $v_0$ is the value of $v$ at the
ingoing ray where the gauge transformation is carried out), in which case
$f_{new}$ turns out to be approximately
constant throughout the ingoing ray. (Another convenient choice is to
{\it define} $u_{new}(u)$ by the demand $f_{new}(u,v_0)=1$.) 
We recall, however, that our goal is to
numerically compute $f_{original}$
(as well as $r$ and $\Phi$) as a function
of $u_{original}$ and $v$, and the gauge transformations are just a
subsidiary manipulation. In order to accomplish this goal,
we must keep record of two additional variables:
(i) the vector $u_{original}(I)$, and (ii) the vector $R_g(I)$,
where $R_g\equiv \,du_{current}/\,du_{original}$ is the cumulative gauge
factor. At each gauge transformation, the latter is updated according to
$R_{g(new)}=(\,du_{new}/\,du_{old}) R_{g(old)}$. Consequently, in the original
gauge the metric function $f$ is given by
$f_{original}=R_g f_{current}$. \footnotemark \footnotetext{
In a point-splitting, the variables
$u_{original}(I)$ and $R_g(I)$ are interpolated at the point added, like the
other variables $h_i(I)$.} Hereafter, whenever we mention $u$ and $f$, we refer to 
$u_{original}$ and $f_{original}$, correspondingly.

The combination of the above four types of numerical manipulations yields
an accurate and efficient numerical code.
It is important to recall that, whereas the
{\it point-splitting} and {\it chopping} are
necessary for an integration to
large values of $v$, the {\it point removal} and {\it gauge correction} are
optional, and are aimed to save integration time\footnotemark \footnotetext{
If a gauge-correction is not used for any reason, then, as a consequence of
successive point-splittings, the difference in $u$ between two adjacent
points near the horizon becomes as small as, say, $10^{-100}$.
Then, due to the roundoff error, it is
not possible to calculate $\,du(I)\equiv u(I+1)-u(I)$ directly at each step.
One way to overcome this difficulty is to keep an independent vector
$\,du(I)$, and to update it at every point-splitting by dividing $\,du(I)$ by
two. (Recall that it is $\,du$ that is involved in the finite-difference
integration scheme, not $u$.) Another
possibility is to shift $u$ by a constant,
e.g. at the end of each ``band,'' so as to assign the AH the
value $u=0$ --- in this case $u(I+1)-u(I)$ can be calculated directly at each
stage.}.
With these two manipulations,
the typical number of grid points in the
vector $u(I)$ grows only logarithmically
with $v_e$, instead of linearly. In practice,
the integration times in long runs
are reduced by a factor of $10$ or so.
Typically, in a running to large $v$,
almost all points in $u$ are
``born'' in point-splitting near the EH, and are later removed when
they approach $r\gg 2M$. We emphasize that in the numerical scheme
described here the increment in $v$ is fixed, $\,dv=1/N$.

\section{Stability, accuracy, and error analysis}
\subsection*{Stability}

Gundlach and Pullin (GP) \cite{GP} recently argued
that any free-evolution scheme will be inherently unstable, in the sense
that small violations of the constraint equations will grow
exponentially with $t$ along lines $r={\rm const}$ --- even if the evolution
equations are exactly satisfied. We disagree with the
theoretical analysis and interpretation made by GP,
for reasons explained in the Appendix.
Also, our numerical tests did not indicate any such numerical instability.
Certain entities exhibit an exponential growth,
but these entities are not the ones that may
be used as authentic error indicators;
rather, the exponential growth we encountered
is merely a reflection of the passive
exponential growth exhibited by various gauge-dependent entities
(e.g., $r_{,u}$, for Kruskal-like $u$) along lines $r={\rm const}$
(or along the EH) in the Schwarzschild geometry.
We discuss this issue extensively in the Appendix.

In our stability tests, we numerically
reconstructed the Schwarzschild spacetime
(as well as RN and other spacetimes) up to $t$ values
of many thousands times $M$, and with values
of the grid-parameter $N$ ranging from 5 to 160. In all these cases,
we found a stable numerical evolution.

\subsection*{Accuracy checks and error analysis}
We used several methods to test and monitor the accuracy of our numerical
code:
(i) Comparing the results obtained with different values of the 
grid-parameter $N$;
(ii) monitoring the discrepancy in the two constraint
equations (\ref {con1},\ref {con2})
(as explained above, our integration scheme
does not involve these equations);
(iii) numerically reproducing known exact solutions:
The Minkowski solution, the vacuum
Schwarzschild solution, the electro-vacuum RN solution,
and the self-similar spherically symmetric scalar-field solution
\cite{roberts} for the Einstein-scalar field equations.
In all these cases our code reproduced the exact solutions very well
(we have compared the metric coefficients $r,f$ and the mass function).
All these tests indicated that the code is stable, and the numerical errors
decrease like $N^{-2}$, as expected from a second-order code (see examples
below).

\begin{figure}
\epsfxsize=8.0cm
\epsffile{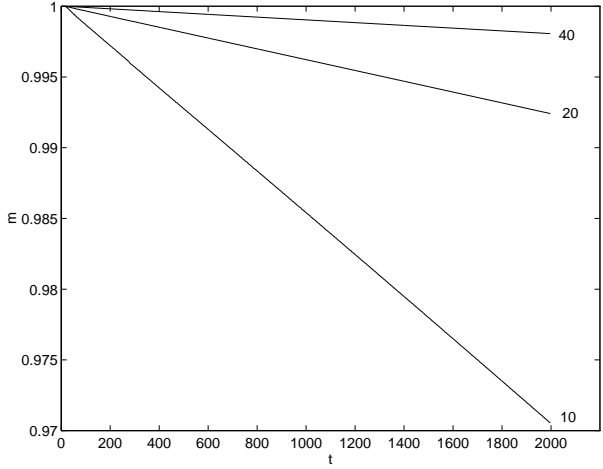}
\caption{The drift of the mass function along $r={\rm const}$. Shown here
is the mass 
function as a function of $t$ for $N=10,20$ and $40$. We took here
$r=8$.}
\label{mast}
\end{figure}

\begin{figure}
\epsfxsize=8.0cm
\epsffile{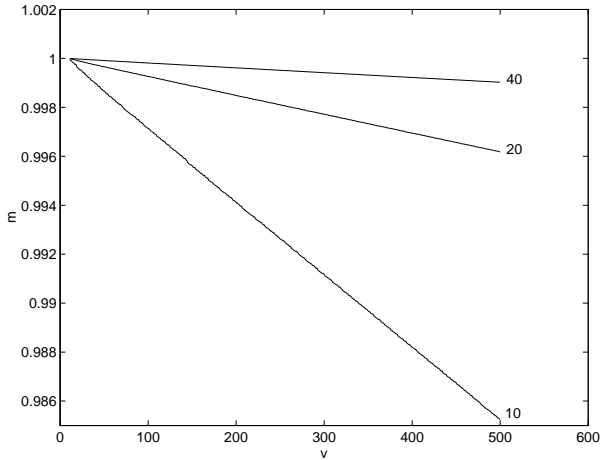}
%
\caption{The drift of the mass function along the EH, as a function of $v$, for 
$N=10,20$ and $40$.}
\label{masv}
\end{figure}

\begin{figure}
\epsfxsize=8.0cm
\epsffile{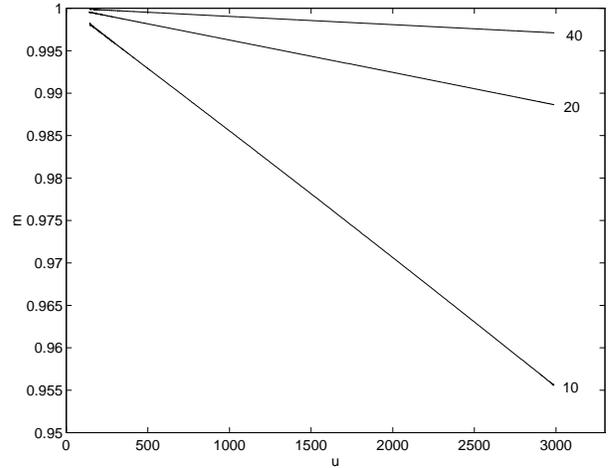}
%
\caption{The drift of the mass function along null infinity, as a function of $u_{e}$ 
(calibrated such that $u_{e}=0$ on $u=0$), for $N=10,20$ and $40$.}
\label{masu}
\end{figure}

We present here the results for the numerical reproduction of the
Schwarzschild solution. (Similar results were obtained for the other 
above-mentioned tests.)  
We start with initial data
corresponding to $M=1$. The drift of $M$ from its
initial value may then be used as an error
indicator. Figure \ref{mast} displays $M$
as a function of $t$ along a line $r={\rm const}$,  Fig.  \ref{masv} 
shows the drift of $M$ as a function of $v$ along the EH, and Fig. \ref{masu} 
shows $M$ as a function of $u_{e}$ along null infinity,  
for various $N$ values. From these Figures it is apparent that the mass 
drift is linear with time, and decreases like $N^{-2}$, as expected. We also 
compared the mass function obtained from local differentiation 
[Eq. (\ref{mass-function})] with 
the dynamical mass function which evolves according to the wave equation  
\cite{Poisson-Israel} 
\begin{equation}
m_{,uv}=2\frac{r^3}{f}\Phi_{,u}^2\Phi_{v}^2-r\;\left(1-\frac{2m}{r}+
\frac{Q^2}{r^2}\right)\;\Phi_{,u}\Phi_{,v}.
\end{equation} 
Both expressions for the mass function agree with each other (in the limit of 
large $N$). 

\begin{figure}
\epsfxsize=8.0cm
\epsffile{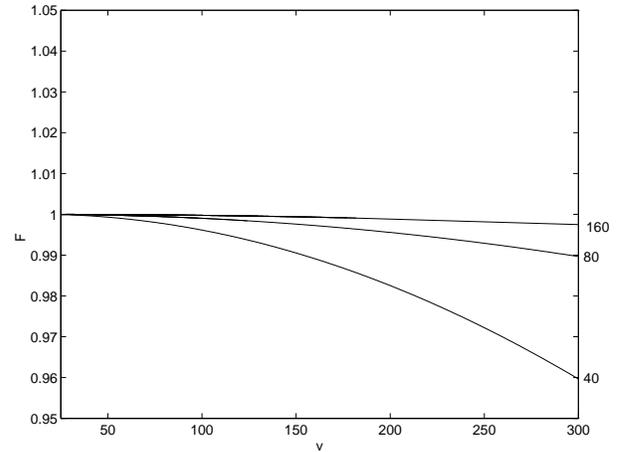}
\caption{The ratio $F$ of $g_{U_kv_e}$ of the exact Schwarzschild 
solution and $g_{U_kv_e}$ of the numerically reproduced 
spacetime. Shown are 
the values for $N=40,80$, and $160$. The convergence indicates a 
second-order code. The deviation of the curves from straight 
lines results primarily from the linear drift of $M$.} 
\label{fig5a} 
\end{figure}

Another check we performed was to compare the values of $f$ as a 
function of $t$ along lines $r={\rm const}$ of the numerically reproduced 
Schwarzschild spacetime and the exact analytical counterpart 
\cite{gundlach-private}. 
Figure \ref{fig5a} 
displays the results we obtained for $r=3M$. (Similar behavior was found 
for other values of $r$.) It is convenient to compare 
the values of $f$ in the  outgoing Kruskal coordinate $U_{k}$ and 
the ingoing Eddington coordinate $v_{e}$. For the Schwarzschild solution 
one finds that $g_{U_{k}v_{e}}=\frac{4M^2}{r}e^{-r/(2M)}
e^{v_{e}/(4M)}$.  
Figure \ref{fig5a} shows the ratio $F$ between $g_{U_{k}v_{e}}$ for the 
exact 
Schwarzschild solution and $g_{U_{k}v_{e}}$ in the numerically reproduced 
spacetime, along $r=3M$, vs. the ingoing coordinate $v$, for several 
values of the grid parameter $N$. This figure clearly indicates  
the second-order convergence of the 
code to the correct theoretical value.

\begin{figure}
\epsfxsize=8.0cm
\epsffile{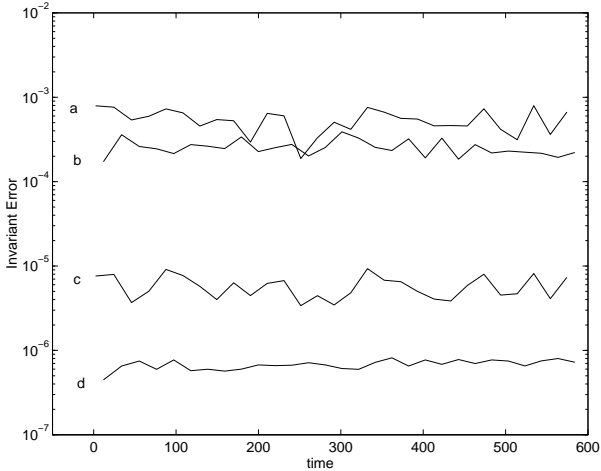}
\caption{The gauge-invariant error indicators  $|C_u|$ and $|C_v|$ 
along a line $r={\rm const}$ as functions of $t$, and 
$|C_u|$ and $|C|$ along the EH as functions of $v_{e}$.  
Line a: $|C_{u}|$ along $r={\rm const}$, line b: $|C_{u}|$ along the EH, 
line c:  $|C_v|$  along $r={\rm const}$, and line d: $|C|$ along the EH. The data 
are taken for $r=3$ and $N=40$.}
\label{c}
\end{figure}

As we mentioned above, we also
use the constraint equations to monitor the errors.
Let us denote the entities in the left-hand side of
Eqs. (\ref {con1},\ref {con2})
by $C'_u$ and $C'_v$, respectively. Now, as they stand, $C'_u$ and $C'_v$
cannot be used as measures of the intrinsic local error in the reproduced
spacetime, because they are not gauge invariant. Instead,
whenever $r_{,u}$ or $r_{,v}$ are non-vanishing, we may define
$$C_u \equiv C'_u/r_{,u}^2 \; , \; C_v \equiv C'_v/r_{,v}^2 \; .$$
Since $C_u$ and $C_v$ are gauge-invariant, they provide an invariant measure
of the local numerical error. An alternative gauge-invariant indicator is
$C\equiv f^{-1}\left(C'_u C'_v\right)^{1/2}$. Note that the
indicator $C_v$ cannot be used at the
AH, where $r_{,v}$ vanishes.
Instead, one may use the indicator $C$ there. Figure \ref{c} displays $C_u$ and
$C_v$ at constant $r$ as functions of $t$ (we took here $r=3$),  
and $C_u$ and $C$ at the EH, as functions of $v$. 
From Figure \ref{c} one can see that these indicators are roughly 
constant with time. (The noise is a results of the second-order numerical 
differentiation necessary for the 
computation of the indicators.) In particular, no exponential growth occurs.
This demonstrates the stability of the code. We found a  
similar behavior also for the electro-vacuum RN spacetime. 

\begin{figure}
\epsfxsize=8.0cm
\epsffile{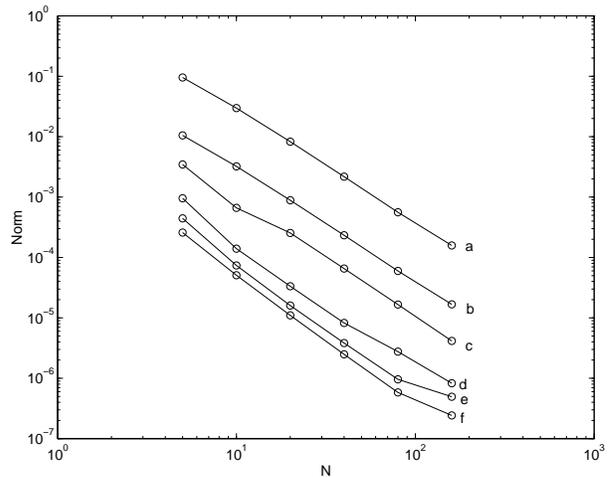}
\caption{The $l_{p}$-norms of the
constraints $C_{u}$ and $C_{v}$ along an 
outgoing null ray, as functions of 
the grid parameter $N$.  
The cases a,b, and c refer to the $l_{1}$,
$l_{2}$, and $l_{\infty}$-norms, respectively, for $C_{u}$, 
and cases d,e, and f refer to the $l_{1}$,
$l_{2}$, and $l_{\infty}$-norms, respectively, for $C_{v}$. The numerical data 
are represented by circles, and the straight lines between the circles  
are linear interpolations of the data.} 
\label{eu}
\end{figure}

Figure  \ref{eu} shows the rate of convergence of various error indicators
as $N$ increases. The spacetime simulated here is RN with
$Q/M=0.8$. (The other exact solutions we checked produced similar results.)
Shown are the $l_{p}$-norms, for several $p$ values,
of the two vectors made of the values of the
indicators $C_{u}$ and $C_{v}$, respectively, along a particular outgoing 
ray located before the EH.    
We used the following values of $N$: 5,10,20,40,80, and 160. The apparently
straight lines in the logarithmic graphs
indicate a second-order convergence\footnotemark\footnotetext{From the 
slopes of the curves displayed in Fig. \ref{eu} we can estimate the 
convergence rate of the code to be around $1.9$, with variations of 
typical order $0.1$. We stress, however, that these numbers would depend 
on the method employed for evaluating the convergence rate.}. (The break in 
the lines e and f for $N=160$ seems             
to be a roundoff effect.) The other error indicators we used (e.g., 
the drift of the mass function and the metric functions) also indicated
a second-order convergence rate.

\section{Non-linear collapse on Minkowski}

In this section we consider
the case $M_0=0,Q=0$, namely, the collapse of the self-gravitating
scalar field over a Minkowski spacetime, leading to the formation of a
Schwarzschild-like black hole. Our initial data correspond to
a compact sinusoidal ingoing pulse, as described in Section II. Here, 
we take $v_1=6 ,v_2=16 ,r_0=6 $, and $r_{u0}=-1/4$
(corresponding to $M_{0}=0$). The final mass of the
black hole is then determined by the pulse amplitude $A$. In what follows we
present the results of a numerical simulation with $A=0.4$, leading to a
final black-hole mass $M_{f}\cong 3.54$. 
(Hereafter, we denote by $M_{f}$ the final mass of the black hole.) 

\begin{figure}
\input epsf 
\centerline{ \epsfysize 7.0cm
\epsfbox{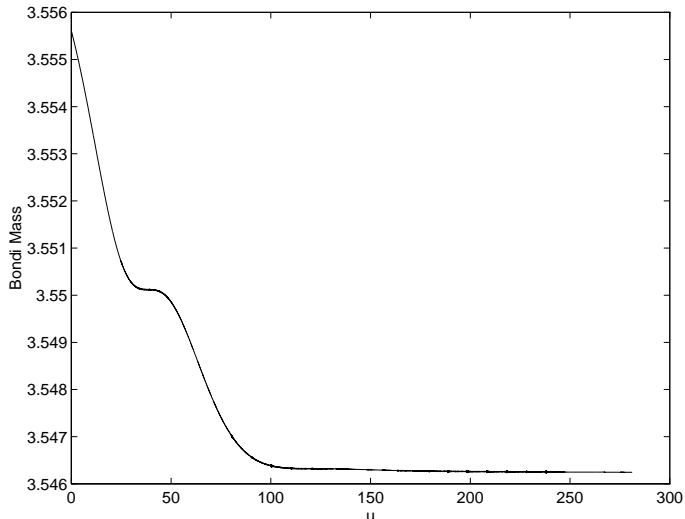}}
%
\caption{The Bondi mass as a function of retarded time $u_e$  
(calibrated such that $u_{e}=0$ on $u=0$), for 
$N=40$. The mass is displayed along the ingoing null ray $v=10^{6}M_{f}$, 
representing null infinity.}
\label{bondi mass}
\end{figure}

Figure \ref{bondi mass} displays the Bondi mass of the
created black hole as a function of the retarded time $u_e$.\footnotemark \footnotetext 
{Strictly speaking, $u_e$ is not well-define here, as the
spacetime is dynamical and differs from Schwarzschild. In the asymptotic
region $r\gg M_{f}$, however,             
the geometry becomes asymptotically Minkowski,                  
and we can define $u_e$ with respect                        
to this asymptotic region. Namely, 
along a ray $v={\rm const}$ in this range, $u_{e}$ is linear with $r$.} 
The Bondi mass decreases with $u_e$,                      
due to the escape of scattered                         
energy to null infinity. The late-time decrease of the mass
corresponding to the power-law tail of the scattered scalar field
is too small to be observed in                                            
this figure (the numerical drift shown in Fig. \ref{masu}        
is also unobservable in the scale used in Fig. \ref{bondi mass}).     
                                                   
\begin{figure}
\input epsf 
\centerline{ \epsfysize 7.0cm
\epsfbox{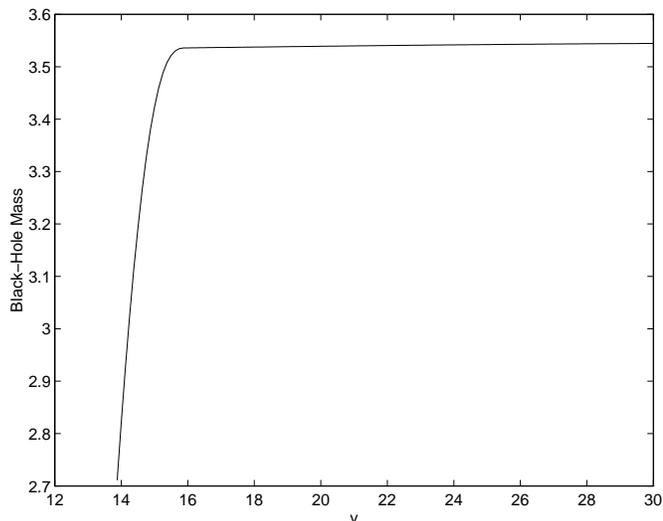}}
%
\caption{Black-hole mass determined from the AH radius vs. $v$.
At advanced times earlier than $v\approx 14$
the domain of integration, $u<u_{f}$, does not intersect the AH.}
\label{ah}
\end{figure}

The nonlinearity of the spacetime dynamics is best represented         
by the evolution of the mass function                                     
along the AH (which is just twice the value of $r$ there).
Figure \ref{ah} shows this mass                                          
as a function of $v$. The mass function                               
grows rapidly, until it approaches a saturation value.                   
In this case, too, the mass-increase at late time due to the power-law
tail, 
and the numerical mass drift, are unseen. The final black-hole mass can be
deduced from either the flat large-$v$ portion of the graph in            
Fig. \ref{ah} or the flat large-$u_e$ portion in      
Fig. \ref{bondi mass}---these two numbers agree, as they should.

\begin{figure}
\input epsf 
\centerline{ \epsfysize 7.0cm
\epsfbox{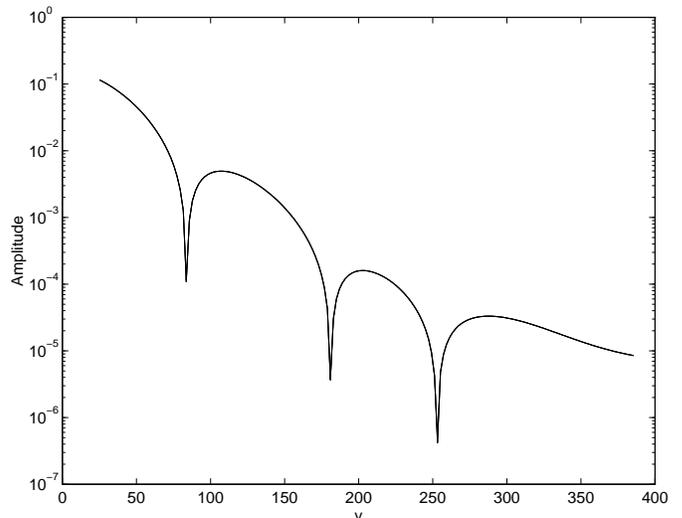}}
%
\caption{Quasi-normal ringing at the horizon as a function of $v_{e}$.  
Recall that $M_{f}\approx 3.54$,
which explains the relatively large value of $v_{e}$ in which
the ringing takes place. 
We used here four different values of $N$---5,10,20, and 40---but 
the four graphs are indistinguishable in this figure.}
\label{stability}
\end{figure}

The stability and accuracy of our code is demonstrated
in Fig. \ref{stability}, which displays the
scalar-field's QN ringing along the horizon for $N=5,10,20$, and $40$: 
The four graphs are indistinguishable in this Figure.  
In addition, we also determined the QN ringing frequency, 
and compared it with the linear 
analysis value \cite{iyer87}. This comparison is hard, as we have only a 
few oscillations before the power-law tails start to dominate. 
In addition, the numerical mass drift (see above) complicates the
comparison between the numerical results and the theoretical prediction.
(However, the mass drift can be controlled by the grid parameter $N$. 
Note that the physical mass increase due to scalar-field absorption is
negligible at late times.) From our numerical data we find that the QN 
frequency is $\sigma=0.032 - 0.026\;i$. 
The real part of $\sigma$ was calculated 
from the two nodes in Fig. \ref{stability}   
corresponding to a full wavelength, and 
the imaginary part from the two local extrema between them. 
The theoretical value for the least damped 
mode with $l=0$ is $\sigma_{th}=0.031 - 0.029\;i$ (recall that here 
$M_f\cong 3.54$).   
The sources for the deviation are the (numerical) drift in the 
mass, the effect of the other $l=0$ modes and the power-law tails, 
and the inability to use values 
from many cycles. However, our numerically obtained value is remarkably 
close to the linear analysis value.

\begin{figure}
\input epsf 
\centerline{ \epsfysize 7.0cm
\epsfbox{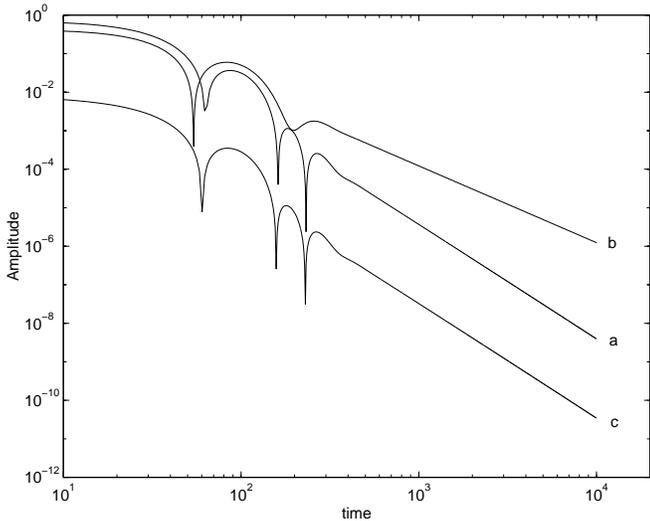}}
%
\caption{The late-time behavior of the scalar field in the three limits
(a),(b), and (c).
Case (a): This graph displays $\Phi$ at $r=2.3M_{f}$, as a function of
$t$.
Case (b): $\Phi$ along null infinity (represented by $v=10^{6}M_{f}$)
versus retarded time $u_e$, calibrated such that $u_{e}=0$ on $u=0$.
Case (c): $\Phi$ along the EH, as a function of $v_{e}$.  
(The amplitude in this case was divided by 100,
so that it will not overlap with the other graphs).
In all three cases, the QN ringing and
the power-law tails are seen clearly. We used here $N=20$.}
\label{amp}
\end{figure}

Figure \ref{amp} shows the late-time behavior of $\Phi$ in the three
asymptotic regions: (a) -- at fixed $r$, with $t\gg M_{f}$ (we 
take $t=(u_e+v_e)/2$), 
(b) -- at future null infinity (represented here by $v_f=10^6 M_{f}$), for
$u_e\gg M_{f}$, and (c) -- at the horizon,
with $v\gg M_{f}$. This figure clearly demonstrates both the QN ringing
and the power-law tails, in all three asymptotic regions.

The determination of the asymptotic behavior at null infinity poses
a special difficulty: We cannot integrate up to $v=\infty$ {\it proper}. (An
attempt to compactify the coordinate $v$ will not solve this problem,
as it would lead to a divergence of
$f$ at null infinity.) We therefore represent null infinity by a large
(buy yet finite)
value, $v=v_f$. This ``null-infinity approximation'' is only valid
as long as $v_f\gg u_e$. Thus, the determination of the late-time
behavior at null infinity clearly demands huge values of $v_f$, in
order to satisfy $v_f\gg u_e\gg M_{f}$. In the simulations described in
this
paper, we used $v_f=10^{6}M_{f}$ to represent null infinity. In
order to enable the integration to such
large $v$ values within a reasonable computation time, we used the following
procedure: Let us denote by $u_{ef}$ the maximal value of $u_e$ in the desired
presentation of the late-time null-infinity behavior (in Fig. \ref{amp},
$u_{ef}=10^4$). After integrating up to a value of $v$ which corresponds 
to $v_{e}=u_{ef}$, we chop the vectors $u(I)$ and 
$h_{i}(I)$ at a value of $u$ which corresponds to $u_{e}=u_{ef}$. The last 
point is now located at $r>2M_f$. 
When we continue the integration to larger $v$ values, the minimal value of $r$,
$r_{min}(v)=r(u_e=u_{ef},v)$, increases very rapidly
and approaches large values (of order $v$). We
can therefore increase $dv$ accordingly, say,
$dv(v)\approx r_{min}(v)/(10N)$. This allows us to integrate up to e.g., 
$v=10^{10}$ within a very short integration time. (Practically, we change
$dv$ in discrete values of $v$, e.g. once in each decade.)

As was mentioned above, one of our goals is to evaluate the
power-law indices in the case of nonlinear collapse and to compare
them to the predictions of the linear theory. Since our initial data
correspond to $l=0$ and to zero initial static moment, the linear
perturbation analysis would predict (negative) power indices 3, 2, and 3
in the cases (a), (b) and (c), correspondingly.
In general, the slopes of the
straight sections in the three graphs shown in
Fig. \ref{amp} appear to agree with these predicted values. However, the
standard best-fit method is not
so useful in this case for a precise determination 
of the numerically-computed indices, due to the following reason.
Consider, for example, the late-time behavior at the horizon. According
to the linear theory, it should be dominated by $v^{-3}$. However, this
dominant term is ``contaminated'' by higher-order terms in $1/v$, whose
effect become larger as $v$ decreases \cite{andersson}. Assume now that we use the
standard best-fit method (applied to a finite interval $v'_0<v<v_f$)
to determine
the deviation of the power-law index from its predicted value. As
it turns out, the computed deviation will be dominated in this case by
the ``higher-order contamination.'' This contamination effect, in turn,
will depend in an arbitrary way on the choice of 
the parameter $v'_0$.
In order to remove this arbitrariness, we introduce the notion of
{\it local power index}, define by $-v \Phi_{,v}/\Phi$. (For the other
asymptotic regions, $v$ is to be replaced by $t$ or $u_e$, accordingly.)

\begin{figure}
\input epsf 
\centerline{ \epsfysize 7.0cm
\epsfbox{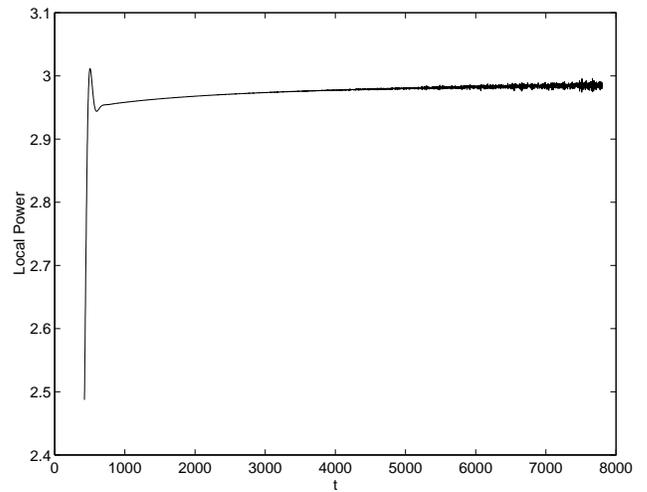}}
\caption{Local index of the `tails' for case (a):
along $r={\rm const}=2.3M_{f}$, as a function of $t$. 
The local power is $2.98\pm 0.01$. We used here $N=20$.}
\label{case a}
\end{figure}

\begin{figure}
\input epsf 
\centerline{ \epsfysize 7.0cm
\epsfbox{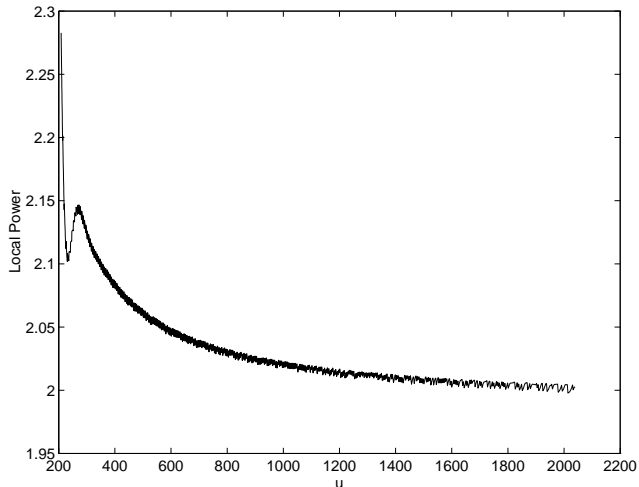}}
\caption{Local index of the `tails' for case (b): along $v=10^{6}M_{f}$
(representing future null infinity), as a function of $u_{e}$. The local 
power is $2.002\pm 0.003$. We used here $N=20$.}
\label{case b}
\end{figure}

\begin{figure}
\input epsf 
\centerline{ \epsfysize 7.0cm
\epsfbox{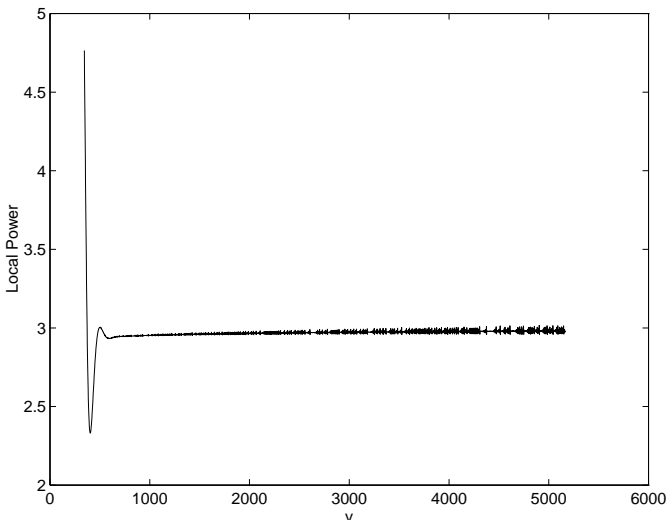}}
%
\caption{Local index of the `tails' for case (c): along the AH, as a 
function of $v_{e}$.
The local power is $2.99\pm 0.02$. We used here $N=20$.}
\label{case c}
\end{figure}

The local power index for the three asymptotic regions is shown in
Figs. \ref{case a}, \ref{case b}, and \ref{case c}. The agreement with the predictions 
of linear theory is
remarkable. In principle, deviations from the precise integer
index may result from three sources of errors:
(i) the limited accuracy of the
numerical simulation; (ii) the finiteness of the late-time domain covered
by the numerics, i.e. the finiteness of $t$, $u$, and $v$ in Figs.
\ref{case a}, \ref{case b}, and \ref{case c}, 
correspondingly (due to the ``higher-orders contamination'',  
the precise integer index is expected only at infinitely-late time);
(iii) in case (b) (i.e. at null infinity), the finiteness of
the final value $v=v_f$ taken to represent null infinity
is also a possible source of error.
In the numerical simulations presented here, we find
that the deviation is related primarily to source (ii): We used a
sufficiently large $N$, and a sufficiently large $v_f$ in case (b), so
sources (i) and (iii) are insignificant.

Our results for the local index (at maximal $t$, $u$, or $v$) are:
\newline
Case (a): $2.98\pm 0.01$ (instead of 3),
\newline
Case (b): $2.002\pm 0.003$ (instead of 2),
\newline
Case (c): $2.99\pm 0.02$ (instead of 3).
\newline
The error bar represents the numerical ``noise'', produced primarily
by the numerical differentiation of $\Phi$ with respect to
$v$, $u$, or $t$,
which is inherent to the computation of the local power index.

For comparison, we quote here the values obtained in previous
nonlinear numerical analyses for the power-law indices.
In case (a): 2.63--2.74 \cite{GPP2} and 3.38 \cite{MC};
in case (c): 3.06 \cite{MC}; No nonlinear results where
obtained so far for case (b).

\section{Non-linear collapse on a charged background}
\begin{figure}
\input epsf 
\centerline{ \epsfysize 7.0cm
\epsfbox{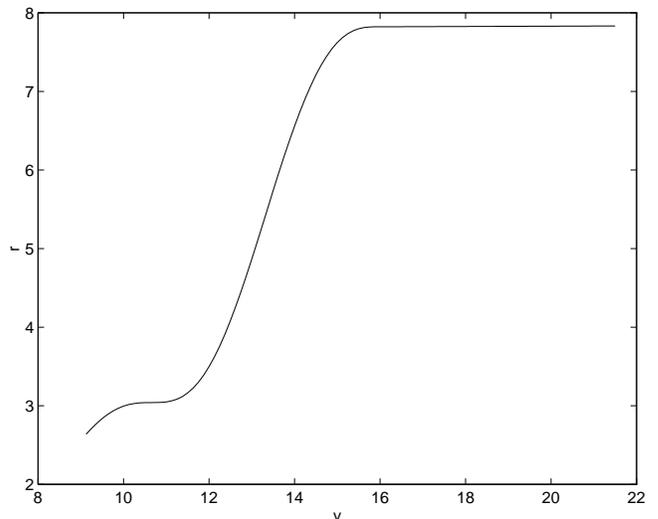}}
%
\caption{Value of $r$ at the AH as a function of $v$.
(At early values of $v$ our numerical domain of
integration $u_i<u<u_f$ does not intersect the AH.) We used 
here $N=20$. }
\label{massincrease}
\end{figure}

In order to study the nonlinear dynamics of charged black holes,
we consider here the
gravitational collapse of the self-gravitating (neutral) scalar field over
a pre-existing charged background (a RN geometry).
The model and initial-value set-up are as explained in Section II.
We take here an initial mass $M_0=1$, and a charge $Q=0.95$.  (We found 
similar results for other values of $Q<1$.) 
We now take $v_1=6 ,v_2=16 ,r_0=6 $, and $r_{u0}\approx -0.1729$.
As before, we take a scalar-field amplitude $A=0.4$. The black-hole mass
then increases to $M_{f}\cong 3.87$ during the collapse.
Figure \ref{massincrease} shows the value of $r$ at the AH vs. $v$.
The rapid increase of the horizon's area indicates strong
nonlinear spacetime dynamics. The two-stage increase of $r$ reflects the 
structure of the scalar-field pulse: The latter has a maximum at about
$v=11$, and the vanishing of $\Phi_{,v}$ implies $M_{,v}=0$ there. (A
similar behavior is observed in Fig. \ref{bondi mass}.) 

\begin{figure}
\input epsf 
\centerline{ \epsfysize 7.0cm
\epsfbox{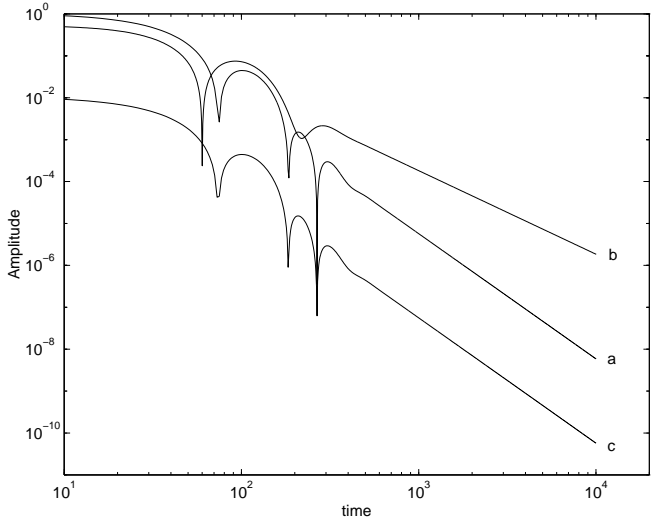}}
%
\caption{Amplitude of the scalar field for the
three cases (a), (b), and (c) for the nonlinear
collapse on a charged RN background. Case (a): along
$r={\rm const}=2.3M_{f}$, as a function of $t$. Case (b): along $v=10^{6}M_{f}$, 
representing future null infinity, as a function of $u_{e}$  
(calibrated such that $u_{e}=0$ on $u=0$).  
Case (c): along the horizon, as a function of $v_{e}$. The amplitude for
case (c) is divided 
by 100 to avoid overlap of the graphs. We used here $N=20$.}
\label{rnt}
\end{figure}

According to the predictions of the linearized
theory \cite{Bicak}, the late-time behavior in the
three asymptotic regions (a,b,c) in a (non-extreme) charged
black hole should be similar to the uncharged case -- namely,
QN ringing followed by inverse power-law tails with the same
indices as in the uncharged case. Figure \ref{rnt}
displays the late-time behavior of the scalar field for the
three asymptotic regions (a), (b), and (c).
Again, both the QN ringing and
the power-law tails are seen very clearly
in all three asymptotic regions.

\begin{figure}
\input epsf 
\centerline{ \epsfysize 7.0cm
\epsfbox{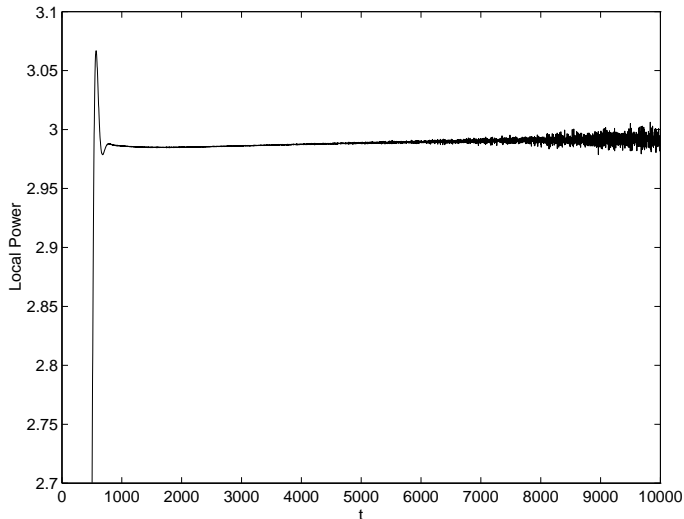}}
%
\caption{Local power as a function of $t$ for case (a):
along constant value of $r$ ($r=2.3M_{f}$). The value
of the local power
approaches $2.99\pm 0.01$. We used here $N=20$.}
\label{rnconst}
\end{figure}

\begin{figure}
\input epsf 
\centerline{ \epsfysize 7.0cm
\epsfbox{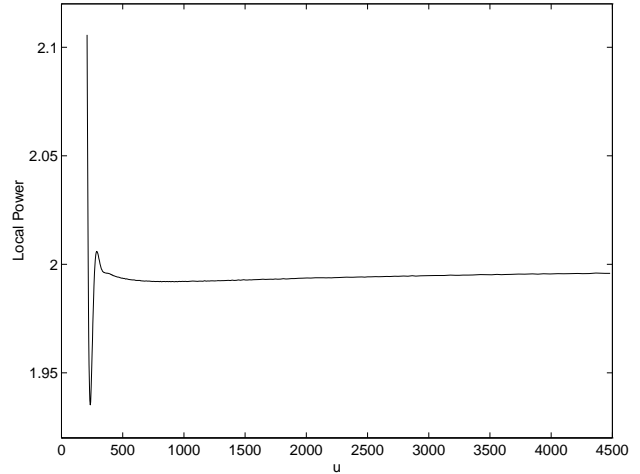}}
%
\caption{Local power as a function of $u_{e}$ for case (b):
along future null infinity, represented by $v=10^{6}M_{f}$.
The value of the local power
approaches $1.996\pm0.001$. We used here $N=20$.}
\label{rnnull}
\end{figure}

\begin{figure}
\input epsf 
\centerline{ \epsfysize 7.0cm
\epsfbox{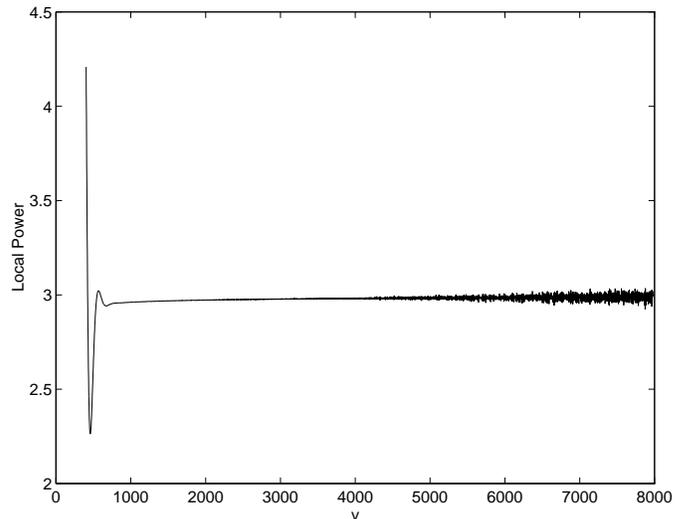}}
%
\caption{Local power as a function of $v_{e}$ for case (c):
along the EH. The value of the local power
approaches $2.99\pm 0.02$. We used here $N=20$.}
\label{rnah}
\end{figure}

Figures \ref{rnconst}, \ref{rnnull}, and \ref{rnah} show the local power 
index for the three asymptotic regions.
Our results for the local power index (at maximal $t$, $u$, or $v$) are:
\newline
Case (a): $2.99\pm 0.01$ (instead of 3),
\newline
Case (b): $1.996\pm 0.001$ (instead of 2),
\newline
Case (c): $2.99\pm 0.02$ (instead of 3).
\newline
These results are in excellent agreement with the predictions
of the linear theory.

\section{Conclusions}
We developed a numerical scheme for the integration of the
spherically-symmetric nonlinear Einstein-Maxwell-Klein-Gordon 
field equations. Our scheme is based on free evolution
in double-null coordinates. This scheme is stable and accurate, it 
is capable of running to arbitrarily late times, and it can handle black 
holes while avoiding singularities. 
  
We used this numerical code to study the gravitational collapse of a
spherically-symmetric, self-gravitating, minimally-coupled 
scalar field to form a black hole (or the collapse of such a scalar field over a pre-existing
charged background). Our numerical simulations demonstrate both the
quasi-normal modes and the power-law tails, in all three late-time
asymptotic regions: at a constant $r$ (with large $t$), along
future null infinity (at large $u$), and along the event horizon (at large $v$). 
The accuracy of our numerical scheme, its ability to run
forever, and the method of calculating {\it local} power indices, 
allowed us to evaluate the power-law indices with an accuracy better 
than all previous estimates.

Our results confirm that the predictions of the linear
theory for the late-time behavior of perturbations outside the black
hole hold also for fully nonlinear collapse.
(This observation is not surprising---in a sense,
it is a manifestation of the principle that black holes have no hair.)
In particular, in all three
late-time asymptotic regions, the power-law indices approach
asymptotically the integer values predicted by
the linear perturbation analysis.
This agreement of the late-time nonlinear dynamics and the linear
perturbation theory was already demonstrated by GPP \cite {GPP2} in the 
uncharged case (see also \cite{andersson}). Here we demonstrate it
for the charged case as well.

The simulations presented here were restricted to the external
part of the black hole and the neighborhood of the event horizon.
One of our main motivations in this project, however, was
to develop the numerical tools which
will allow the investigation of the {\it inner} part of black
holes. We are currently using this numerical scheme to study the
evolution of the geometry and the scalar field near the
spacetime singularity inside a charged black hole.

\section*{Acknowledgment}
This research was supported in part by the Israeli Science Foundation 
administered by the Israel Academy of Sciences and Humanities.

\section*{Appendix A}
At issue here is the evolution of the entities at the
left-hand sides of Eqs. (\ref {con1},\ref {con2}), which represent
the violation of the constraint equations. These entities are denoted in
Ref. \cite{GP} by $E_1$ and $E_2$, respectively
(in Section IV we denote these entities by $C'_u$ and $C'_v$).
GP argue that, as a consequence of the {\it precise} evolution equations,
$E_1$ and $E_2$ will grow exponentially with $t$ along lines $r={\rm const}$.
According to GP, this exponential divergence represents an inevitable
numerical instability of the free-evolution scheme. We do not accept this
conclusion, and claim that
(i) under the precise evolution equations $E_1$ and $E_2$
do {\it not} grow exponentially; rather, they are essentially conserved
(in a sense which will be explained below);
(ii) in our numerical free-evolution scheme,
$E_1$ (but not $E_2$) will grow exponentially,
but this growth is a consequence of the {\it numerical error}
in the integration of the evolution equations, not of
the equations themselves. 
(iii) This divergence of $E_1$ does not
indicate a numerical instability; Rather, it reflects
the passive exponential growth of typical gauge-dependent entities
like $r_{,u}$ along lines $r={\rm const}$ in the Schwarzschild geometry. 

To verify point (i),
assume that the evolution equations are precisely satisfied. Then,
Eq. (7) in Ref. \cite{GP} reads
$E_{1,v}=-(r_{,v}/r) E_1$. $E_2$ will satisfy an analogous equation:
$E_{2,u}=-(r_{,u}/r) E_2$. It then follows that the
entity $r E_1$ is conserved along lines $u={\rm const}$, and similarly
$r E_2$ is conserved along lines $v={\rm const}$.
Therefore, an exponential growth of $E_1$ or $E_2$
along lines $r={\rm const}$ is ruled out\footnotemark \footnotetext{
Such an exponential divergence would demand that, on the
initial hypersurface, $E_1$ diverges exponentially with $u_e$, and
$E_2$ diverges (almost) exponentially with $v_e$, which is 
an unreasonable situation.}. 
The exponential divergence of the linear metric perturbations (denoted $\xi$ 
and $\eta$ in Ref. \cite{GP}) found by GP must therefore be a gauge 
mode. In fact, the infinitesimal coordinate transformation $u\to u+\,du$
(e.g., with fixed infinitesimal $\,du$) yields a nonvanishing $\xi$ (in
Ref. \cite{GP}  $\xi$ represents the linear
perturbation in $r^2$), given by
$$\xi =-(r^2)_{,u}\,du=-2r\,r_{,u}\,du\;.$$
For any Kruskal-like $u$, at a fixed $r$,
$r_{,u}$ grows like $\exp(t/4M)$ at $t\gg M$. It then follows that
along any line $r={\rm const}$, at large $t$, $\xi$ will
exhibit this $\exp(t/4M)$ divergence.
This fits very well with the rate of divergence, $\exp(0.24\,t/M)$, found
numerically by GP. But this is, of course, a gauge mode,
which does not indicate a violation of the Einstein equations.

Note also that the analytic derivation of the exponential growth in Ref.
\cite{GP} is based on the ``mode ansatz'' approximation. This approximation
may only be valid if the mode's wave-number $k$ is sufficiently large. The
diverging modes found by GP do {\it not} satisfy this
condition, however.  This may explain the discrepancy of the value 
$0.32\;M^{-1}$ predicted by GP 
compared with the above theoretical value $1/(4M)$ (which is also 
confirmed numerically by GP to a good accuracy). 

In our numerical tests, we found that along lines $r={\rm const}$, $E_2$ is
roughly preserved, while $E_1$ grows like $\exp(t/2M)$. From the above
discussion it is obvious that this exponential growth of $E_1$ must be a
consequence of the {\it violation} of
the evolution equations, due to numerical
errors. Later we shall give a more explicit explanation for this
behavior. The crucial point is, however, that the exponential
growth of $E_1$ does not indicate an exponential growth of the intrinsic
local error: The entity $E_1$ (like $E_2$) is not an
appropriate error indicator, because it
is not a gauge-invariant entity. [In a transformation $u\to u'(u)$, $E_1$
is multiplied by the factor $(\,du/\,du')^2$.] In order to extract from $E_1$
the information about the intrinsic local error, we must
construct a gauge-invariant entity from it. A convenient choice is the
gauge-invariant entity $C_u\equiv E_1/(r_{,u})^2$. This entity
indeed remains roughly constant along lines $r={\rm const}$ and along
the EH (see Fig. \ref{c}). The behavior of the other error
indicators (e.g. the mass parameter in Figs. \ref{mast} and \ref{masv}) also
indicate stability: None of the invariant entities exhibit an
exponential growth of error.

We still need to explain why the numerical errors in the integration of
the evolution equations results in an exponential growth of $E_1$.
If there were no numerical errors in the integration, then,
along a line $r={\rm const}\equiv r'$, $E_1$ would approach (at large $t$) a
constant value, $E_1(u_h,v_i) \, r(u_h,v_i)/r'$. Correspondingly, $C_u$
would decay like $1/(r_{,u})^2\propto \exp(-t/2M)$.
However, the numerical error in the integration of
the evolution equations provide a (roughly) constant source
of error in the evolution of $C_u$\footnotemark \footnotetext{
To understand why this source of error is roughly
independent of $t$, recall that (i) in our integration scheme, due
to the dynamical grid refinement, the intrinsic error production rate is
essentially independent of $t$ (or $u$), and (ii)
$C_u$ is an authentic indicator of
the intrinsic local error.}. 
The combination of the dynamical tendency to exponential decay
and the constant source of error
results in a finite saturation value (proportional to the
numerical error)\footnotemark \footnotetext{
Phenomenologically, we may represent the situation by the simple
differential equation $\,dC_u/\,dt=-C_u /(2M)+S(r)$, where $S(r)$ is
the numerical source term. $C_u$ then approaches the asymptotic value
$2MS(r)$.}. 
In turn, this implies that $E_1$ grows like $\exp(t/2M)$.

Finally, we emphasize again that despite of the above discussion,
whenever the domain of integration includes the EH, 
a naive attempt to use a free-evolution scheme in
double-null coordinates, without a grid
refinement, will inevitably result in a numerical instability
due to the exponential divergence of $r_{,u}$
(see Section III). This might be the reason for the failure
of previous attempts to use the free-evolution scheme.
The grid refinement (in our case, the point-splitting procedure)
is a necessary ingredient in such a numerical scheme.


\begin{references}
\bibitem{gundlach-private} This check was suggested to us by C. Gundlach, private 
communication.
\end{references}
\end{document}